\documentclass[preprint]{aastex}

\shorttitle{Comments} \shortauthors{H.Tong}

\begin{document}

\title{Comments on ``No magnetars in ULXs''}

\author{H. Tong (Guangzhou University)}


\begin{abstract}
The magnetic dipole field of ultraluminous X-ray pulsars may not be very high. However, it is too early to say that they
are not magnetars. The existence of low magnetic field magnetars should be taken into consideration.
\end{abstract}


\section{Comments}

King \& Lasota (2019) claimed that there are ``No magnetars in ULXs''. In their opinion, ULX pulsars may
be beamed due to super-Eddington mass transfer. While the neutron star is still accreting matter near the Eddington threshold. Using the
spin-up measurements, the magnetic dipole field is found to be of normal value ($\sim 10^{12} \ \rm G$). Then they claim that there are no magnetars
in ULXs. However, this is just one solution to the ULX pulsar problem. The super-Eddington luminosity of ULX pulsar may be due to the presence of
magnetar strength magnetic field ($\sim 10^{14} \ \rm G$) in the vicinity of the neutron star\footnote{This is also the key point to propose the existence of magnetars in the early 1990s (Duncan \& Thompson 1992; Paczynski 1992; Usov 1992). Therefore, seeing a super-Eddington luminosity, proposing the idea of accreting magnetar is similar to the invention of magnetars. It is not a ``magnetic analogy'' to intermediate mass black holes.} (Paczynski 1992; Mushtukov et al. 2015). The large scale magnetic dipole field of ULX pulsars may be of normal value (Tong 2015; Dall'Osso et al. 2015; Israel et al. 2017b). Therefore, ULX pulsars can also be accreting low magnetic field magnetars.

In addition to the luminosity and spin-up observations, ULX pulsars also have pulse profile and pulsed fraction measurements (which are unique to pulsars).
For the four ULX pulsars (M82 X-2, Bachetti et al. 2014; NGC 7793 P13, F$\rm \ddot{u}$rst et al. 2016, Israel et al. 2017a; NGC 5907 ULX, Israel et al. 2017b; NGC 300 ULX1, Carpano et al. 2018), they all have near sinusoidal pulse profiles. This means that they do not have a strong beaming. Their pulsed fraction are relatively high (e.g., $20\%$ for NGC 5907 ULX, Israel et al. 2017b). Therefore, for the pulsed component alone, ULX pulsars may be accreting at super-Eddington rate. This is in contradiction with that of King \& Lasota (2019). Previous studies already found that the magnetic dipole field of ULX pulsars may not be very high (Tong 2015; Dall'Osso et al. 2015; Israel et al. 2017b). Noting this point, one way is to propose the idea of accreting low magnetic field magnetars (Tong 2015; Israel et al. 2017b). Another way is to consider the ``no magnetar'' solution as in King \& Lasota (2019). One reason for the later choice may be that: the idea of low magnetic field magnetar is not fully appreciated, especially outside the magnetar domain (Rea et al. 2010).

Besides normal magnetars and low magnetic field magnetars, the existence of high magnetic field pulsars (Ng \& Kaspi 2010) further complicates this problem. High magnetic field pulsars are neutron stars with strong magnetic dipole field. But it is not granted they are magnetars. Magnetar is not simply a neutron star with high magnetic dipole field. However, this definition of magnetars are often taken as granted in the accreting neutron star and gamma-ray burst studies. Therefore, seeing a strong magnetic dipole field in slow pulsation X-ray pulsars does not grant their magnetar nature (Sanjurjo-Ferrrin et al. 2017). Similarly, a low magnetic dipole field in ULX pulsars can not rule out the presence of magnetars. At present, they are all ``accreting magnetar candidates'', which are often shortened as ``accreting magnetars''.

In summary, as a new specimen to the zoo of accreting neutron stars, the idea of accreting magnetars should be welcomed and explored in full detail in the future.







\end{document}